\documentstyle[11pt,ska,twoside,psfig]{article}
\begin{document}
\title{Large-scale Structure with the SKA}
 \author{Peter Coles}
\affil{School of Physics \& Astronomy, University of Nottingham,
University Park, Nottingham NG7 2RD, United Kingdom}


\begin{abstract}
A standard theoretical paradigm for the formation of large-scale
structure in the distribution of galaxies has now been
established, based on the gravitational instability of cold dark
matter in a background cosmology dominated by vacuum energy.
Significant uncertainties remain in the modelling of complex
astrophysical processes involved in galaxy formation, perhaps most
fundamentally in the relationship between the distributions of
luminous galaxies and the underlying dark matter. I argue that the
Square Kilometre Array is likely to provide information crucial to
understanding this relationship and how it evolves with time.
\end{abstract}

\section{Introduction}

Over the last few years, cosmology has witnessed unprecedented
improvements in our knowledge of the basic parameters governing
the expansion of the Universe originating with observations of
high-redshift supernovae (Riess et al. 1998; Perlmutter et al.
1999) and culminating in the recently-released data from the WMAP
satellite (Spergel et al. 2003). As a result of these developments
and parallel advances in theory, the field of large-scale
structure has now entered a period of transition. Before the onset
of the current data explosion, there were two basic reasons for
wanting to study galaxy clustering. One was that it might furnish
observational ways of pinning down cosmological parameters, and
the other was that it provided the context within which to study
galaxy formation and evolution. These two approaches are not
mutually exclusive, of course, but one might associate the first
with cosmologists of a more astrophysical persuasion, whereas the
second is more likely to come from particle-cosmologists or
inflationary specialists.  Both points of view have stimulated the
development of this field over the past twenty years. Now things
are changing. Given the apparent precision with which we now know
the cosmological parameters, observations of galaxy clustering
will at most be seen as consistency checks on the fundamental
properties of the Universe. On the other hand, accelerating
improvements of observational technology have opened up the
possibility of probing the very detailed and subtle properties of
galaxies that are regarded as a nuisance to those interested in
fundamental parameters.

In this paper, I will argue that the strongest contribution likely
to be made by Square Kilometre Array, given the timescale required
for its completion, is likely not to be in the pristine world of
particle cosmology but in the grubby astrophysics of galaxy
formation. I start by giving a very brief overview of structure
formation theory for non-specialists and then try to draw out some
of the areas in which the character of the subject is changing. I
will then discuss briefly the merits of 21cm galaxy surveys
discussed in the science case for SKA which can be found at:

 \begin{center} {\tt \begin{verbatim}
http://www.skatelescope.org/ska_science.shtml \end{verbatim}}
\end{center}

\section{Basics of Cosmological Structure Formation}

\subsection{Basic Framework}

The Big Bang theory is built upon the Cosmological Principle,
which requires the Universe on large scales to be both homogeneous
and isotropic. Space-times consistent with this requirement can be
described by the Robertson--Walker metric
\begin{equation}
{\rm d}s_{\rm FRW}^2 = c^2 {\rm d}t^2 - a^2(t)\left({{\rm
d}r^2\over 1 - \kappa r^2} + r^2 {\rm d}\theta^2 + r^2\sin^2\theta
{\rm d}\phi^2\right)  , \label{eq:l1a}
\end{equation}
where $\kappa$ is the spatial curvature, scaled so as to take the
values $0$ or $\pm 1$. The case $\kappa=0$ represents  flat space
sections, and the other two cases are  space sections of constant
positive or negative curvature, respectively. The time coordinate
$t$ is called {\em cosmological proper time} and it is singled out
as a preferred time coordinate by the property of spatial
homogeneity. The quantity $a(t)$,  the {\em cosmic scale factor},
describes the overall expansion of the universe  as a function of
time. If light emitted at time $t_{\rm e}$ is received by an
observer at $t_0$ then the redshift $z$ of the source is given by
\begin{equation}
1+z = \frac{a(t_0)}{a(t_{\rm e})}.
\end{equation}
The dynamics of an FRW universe are determined by the Einstein
gravitational field equations which become
\begin{eqnarray}
3\left( \frac{\dot{a}}{a} \right)^{2} & = & 8\pi G\rho - {3\kappa
c^{2} \over a^2} + \Lambda,\\ {\ddot{a}\over a} & = & - {4\pi
G\over 3} \left(\rho + 3 \frac{p}{c^2}\right) + {\Lambda\over 3},
\\ \dot{\rho}& =& - 3 {\dot{a}\over a}\left(\rho + \frac{p}{c^2}
\right). \label{eq:l1b}
\end{eqnarray}
These equations  determine the time evolution of the cosmic scale
factor $a(t)$ (the dots denote derivatives with respect to
cosmological proper time $t$) and therefore describe the global
expansion or contraction of the universe. The behaviour of these
models can further be parametrised in terms of the Hubble
parameter $H=\dot{a}/a$ and the density parameter $\Omega=8\pi
G\rho/3H^2$, a suffix $0$ representing the value of these
quantities at the present epoch when $t=t_0$. The cosmological
constant is denoted $\Lambda$ here, but it can be regarded instead
as an additional energy density various forms of which have a
similar effect; see Huterer \& Turner (2001).

\subsection{Linear Theory}

In order to understand how structures form we need to consider the
difficult problem of dealing with the evolution of inhomogeneities
in the expanding Universe . We are helped in this task by the fact
that we expect such inhomogeneities to be of very small amplitude
early on so we can adopt a kind of perturbative approach, at least
for the early stages of the problem. If the length scale of the
perturbations is smaller than the effective cosmological horizon
$d_H=c/H_0$, a Newtonian treatment of the subject  is expected to
be valid.  If the mean free path of a particle is small, matter
can be treated as an ideal fluid and the Newtonian equations
governing the motion of gravitating particles in an expanding
universe can be written in terms of ${\bf x} = {\bf r} / a$ (the
comoving spatial coordinate, which is fixed for observers moving
with the Hubble expansion), ${\bf v} = \dot {{\bf r}} - H {\bf r}
= a\dot {{\bf x}}$ (the peculiar velocity field, representing
departures of the matter motion from pure Hubble expansion), $\phi
({\bf x} , t)$ (the peculiar Newtonian gravitational potential,
i.e. the fluctuations in potential with respect to the homogeneous
background) and $\rho ({\bf x}, t)$ (the matter density). Using
these variables
 we obtain, first, {\em the Euler equation}:
\begin{equation}
{\partial (a{\bf v})\over \partial t} + ({\bf v}\cdot{\bf
\nabla_x}){\bf v} = - {1\over \rho}{\bf \nabla_x} p - {\bf
\nabla_x}\phi~. \label{eq:Euler}
\end{equation}
The second term on the right-hand side of equation (6) is the
peculiar gravitational force, which can be written in terms of
${\bf g} = -{\bf \nabla_x}\phi/a$, the peculiar gravitational
acceleration of the fluid element. If the velocity flow is
irrotational, ${\bf v}$ can be rewritten in terms of a velocity
potential $\phi_v$: ${\bf v} = - {\bf \nabla_x} \phi_v/a$. Next we
have the {\em continuity equation}:
\begin{equation}
{\partial\rho\over \partial t} + 3H\rho + {1\over a} {\bf
\nabla_x} (\rho{\bf v}) = 0, \label{eq:continuity}
\end{equation}
which expresses the conservation of matter, and finally the {\em
Poisson equation}:
\begin{equation}
{\bf \nabla_x}^2\phi = 4\pi G a^2(\rho - \rho_0) = 4\pi
Ga^2\rho_0\delta, \label{eq:Poisson}
\end{equation}
describing Newtonian gravity.  Here $\rho_0$ is the mean
background density, and
\begin{equation}
\delta \equiv \frac{\rho-\rho_0}{\rho_0}
\end{equation}
is the {\em density contrast}.

The next step is  to linearise the Euler, continuity and Poisson
equations by perturbing physical quantities defined as functions
of Eulerian coordinates, i.e. relative to an unperturbed
coordinate system. Expanding $\rho$, ${\bf v}$ and $\phi$
perturbatively and keeping only the first-order terms in equation
(7) gives the linearised continuity equation:
\begin{equation}
{\partial\delta\over \partial t} = - {1\over a}{\bf \nabla_x}\cdot
{\bf v},
\end{equation}
which can be inverted, with a suitable choice of boundary
conditions, to yield
\begin{equation}
\delta = - {1\over a H f}\left({\bf \nabla_x}\cdot{\bf v}\right).
\label{eq:l9}
\end{equation}
The function $f\simeq \Omega_0^{0.6}$; this is simply a fitting
formula to the full solution. The linearised Euler and Poisson
equations are
\begin{equation}
{\partial {\bf v}\over\partial t} + {\dot a\over a}{\bf v} = -
{1\over \rho a}{\bf \nabla_x} p -{1\over a}{\bf \nabla_x}\phi,
\label{eq:l10}
\end{equation}
\begin{equation}
{\bf \nabla_x}^2\phi = 4\pi G a^2\rho_0\delta; \label{eq:l11}
\end{equation}
$|{\bf v}|, |\phi|, |\delta| \ll 1$ in equations (10), (12) \&
(13). From these equations, and if one ignores pressure forces, it
is easy to obtain an equation for the evolution of $\delta$:
\begin{equation}
\ddot\delta + 2H\dot\delta - {3\over 2}\Omega H^2\delta = 0.
\label{eq:l13b}
\end{equation}
For a spatially flat universe dominated by pressureless matter,
$\rho_0(t) = 1/6\pi Gt^2$ and equation (14) admits two linearly
independent power law solutions $\delta({\bf x},t) =
D_{\pm}(t)\delta({\bf x})$, where $D_+(t) \propto a(t) \propto
t^{2/3}$  is the growing  mode and $D_-(t) \propto t^{-1}$ is the
decaying mode.

\subsection{Primordial density fluctuations}
The above considerations apply to the evolution of a single
Fourier mode of the density field $\delta({\bf x}, t) =
D_+(t)\delta({\bf x})$. What is more likely to be relevant,
however, is the case of a superposition of waves, resulting from
some kind of stochastic process in which he density field consists
of a  superposition of such modes with different amplitudes. A
statistical description of the initial perturbations is therefore
required, and any comparison between theory and observations will
also have to be statistical.

The spatial Fourier transform of $\delta({\bf x})$ is
\begin{equation}
\hat{\delta}({\bf k}) = \frac{1}{(2\pi)^{3}} \int {\rm d}^{3} {\bf
x}e^{- i{\bf k}\cdot{\bf x}} \delta({\bf x}). \label{eq:lad1}
\end{equation}
It is useful to specify the properties of $\delta$ in terms of
$\hat{\delta}$. We can define the {\em power-spectrum} of the
field to be (essentially) the variance of the amplitudes at a
given value of ${\bf k}$:
\begin{equation}
\langle \hat{\delta}({\bf k}_1) \hat{\delta}({\bf k}_2) \rangle =
P(k_1) \delta^{D} ({\bf k}_1+{\bf k}_2), \label{eq:lad4}
\end{equation}
where $\delta^{D}$ is the Dirac delta function; this rather
cumbersome definition takes account of the translation symmetry
and reality requirements for $P(k)$; isotropy is expressed by
$P({\bf k})=P(k)$. The analogous quantity in real space is called
the two-point correlation function or, more correctly, the
autocovariance function, of $\delta({\bf x})$:
\begin{equation}
\langle \delta({\bf x}_1) \delta({\bf x}_2) \rangle =\xi (|{\bf
x}_1-{\bf x_2}|) = \xi({\bf r})=\xi(r), \label{eq:lad5}
\end{equation}
which is itself related to the power spectrum via a Fourier
transform. The power-spectrum is particularly important because it
provides a complete statistical characterisation of a particular
kind of stochastic process: a {\em Gaussian random field}. This
class of field is the generic prediction of inflationary models,
in which the density perturbations are generated by Gaussian
quantum fluctuations  in a scalar field during the inflationary
epoch (e.g. Brandenberger 1985).

The shape of the initial fluctuation spectrum, is assumed to be
imprinted on the universe at some arbitrarily early time. Many
versions of the inflationary scenario for the very early universe
(Guth 1981) produce  a power-law form
\begin{equation}
P(k)=Ak^{n}, \label{eq:lad10}
\end{equation}
with a preference in some cases for the Harrison--Zel'dovich form
with $n=1$ (Harrison 1970; Zel'dovich 1972). Even if inflation is
not the origin of density fluctuations, the form (18) is a useful
phenomenological model for the fluctuation spectrum.

These considerations specify the shape of the fluctuation
spectrum, but not its amplitude. The discovery of temperature
fluctuations in the CMB (Smoot et al. 1992) plugged that gap.

\subsection{The transfer function}
We have hitherto assumed that the effects of pressure and other
astrophysical processes on the gravitational evolution of
perturbations are negligible. In fact, depending on the form of
any dark matter, and the parameters of the background cosmology,
the growth of perturbations on particular length scales can be
suppressed relative to the growth laws discussed above.

We need first to specify the fluctuation mode. In cosmology, the
two relevant alternatives are {\em adiabatic} and {\em
isocurvature}. The former involve coupled fluctuations in the
matter and radiation component in such a way that the entropy does
not vary spatially; the latter have zero net fluctuation in the
energy density and involve entropy fluctuations. Adiabatic
fluctuations are the generic prediction from inflation and form
the basis of most currently fashionable models, although
interesting work has been done  on isocurvature models (e.g.
Peebles 1999).

In the classical Jeans instability, pressure inhibits the growth
of structure on scales smaller than the distance traversed by an
acoustic wave during the free-fall collapse time of a
perturbation. If there are collisionless particles of hot dark
matter, they can travel rapidly through the background and this
free streaming can damp away perturbations completely. Radiation
and relativistic particles may also cause kinematic suppression of
growth. The imperfect coupling of photons and baryons can also
cause dissipation of perturbations in the baryonic component. The
net effect of these processes, for the case of statistically
homogeneous initial Gaussian fluctuations, is to change the shape
of the original power-spectrum in a manner described by a simple
function of wave-number -- the transfer function $T(k)$ -- which
relates the processed power-spectrum $P(k)$ to its primordial form
$P_0(k)$ via $P(k) = P_0(k)\times T^2(k)$. The results of full
numerical calculations of all the physical processes we have
discussed can be encoded in the transfer function of a particular
model (Bardeen et al. 1986). For example, fast moving or `hot'
dark matter particles (HDM) erase structure on small scales by the
free-streaming effects mentioned above so that $T(k)\rightarrow 0$
exponentially for large $k$; slow moving or `cold' dark matter
(CDM) does not suffer such strong dissipation, but there is a
kinematic suppression of growth on small scales (to be more
precise, on scales less than the horizon size at matter--radiation
equality); significant small-scale power nevertheless survives in
the latter case. These two alternatives thus furnish two very
different scenarios for the late stages of structure formation:
the `top--down' picture exemplified by HDM first produces
superclusters, which subsequently fragment to form galaxies; CDM
is a `bottom--up' model because small-scale structures form first
and then merge to form larger ones. The general picture that
emerges is that, while the amplitude of each Fourier mode remains
small, i.e. $\delta({\bf k})\ll 1$, linear theory applies. In this
regime, each Fourier mode evolves independently and the
power-spectrum therefore just scales as
\begin{equation}
P(k,t)=P(k,t_1) {D_{+}^{2}(k,t)\over D_{+}^2(k,t_1)} = P_0(k)
T^{2} (k) {D_{+}^{2}(k,t)\over D_{+}^2(k,t_1)}~.
\end{equation}
For scales larger than the Jeans length, this means that the shape
of the power-spectrum is preserved during linear evolution.

\begin{figure}
\begin{center}
\centerline{\psfig{figure=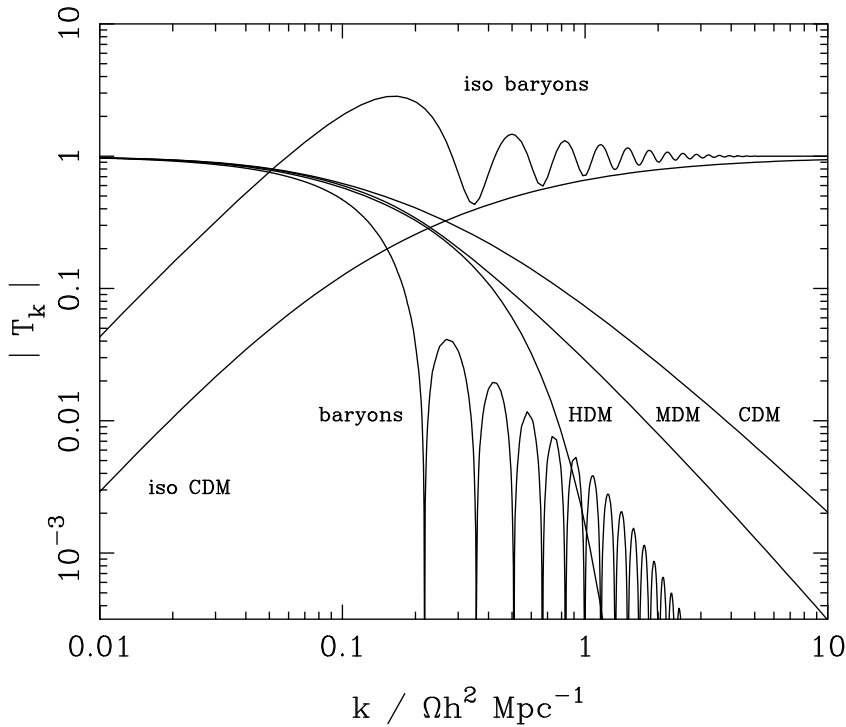,height=10cm,angle=0}}
\caption{Examples of adiabatic transfer functions for baryons, hot
dark matter (HDM), cold dark matter (CDM) and mixed dark matter
(MDM; also known as CHDM). Isocurvature modes are also shown.
Picture courtesy of John Peacock.}
\end{center}
\end{figure}

\subsection{Beyond linear theory}
The linearised equations of motion  provide an excellent
description of gravitational instability at very early times when
density fluctuations are still small ($\delta \ll 1$). The linear
regime of gravitational instability breaks down when $\delta$
becomes comparable to unity, marking the commencement of the {\it
quasi-linear} (or weakly non-linear) regime. During this regime
the density contrast may remain small ($\delta < 1$), but the
phases of the Fourier components $\delta_{\bf k}$ become
substantially different from their initial values resulting in the
gradual development of a non-Gaussian distribution function if the
primordial density field was Gaussian. In this regime the shape of
the power-spectrum changes by virtue of a complicated cross-talk
between different wave-modes. Analytic methods are available for
this kind of problem , but the usual approach is to use $N$-body
experiments for strongly non-linear analyses (Davis et al. 1985;
Jenkins et al. 1999).

Further into the non-linear regime, bound structures form. The
baryonic content of these objects may then become important
dynamically: hydrodynamical effects (e.g. shocks), star formation
and heating and cooling of gas all come into play. The spatial
distribution of galaxies may therefore be very different from the
distribution of the (dark) matter, even on large scales. Attempts
are only just being made to model some of these processes with
cosmological hydrodynamics codes, but it is some measure of the
difficulty of understanding the formation of galaxies and clusters
that most studies have only just begun to attempt to include
modelling the detailed physics of galaxy formation. In the front
rank of theoretical efforts in this area are the so-called
semi-analytical models which encode simple rules for the formation
of stars within a framework of merger trees that allows the
hierarchical nature of gravitational instability to be explicitly
taken into account (Baugh et al. 1998).

The usual approach is instead simply to assume that the point-like
distribution of galaxies, galaxy clusters or whatever,
\begin{equation}
n({\bf r})=\sum_i\delta_D({\bf r} - {\bf r}_i),
\end{equation}
bears a simple functional relationship to the underlying
$\delta({\bf r})$.
 An assumption often invoked is that relative fluctuations in the object number
counts and matter density fluctuations are proportional to each
other, at least within sufficiently large volumes, according to
the {\em linear biasing} prescription:
\begin{equation}
{\delta n({\bf r})\over \bar n}~=~b\,{\delta \rho({\bf r})\over
\bar \rho}\,, \label{eq:linb}
\end{equation}
where $b$ is what is usually called the biasing parameter.
Alternatives, which are not equivalent, include the high-peak
model (Kaiser 1984; Bardeen et al. 1986) and the various local
bias models (Coles 1993). Non-local biases are possible, but it is
rather harder to construct such models (Bower et al. 1993).  If
one is prepared to accept an {\it ansatz} of the form (21) then
one can use linear theory on large scales to relate galaxy
clustering statistics to those of the density fluctuations, e.g.
\begin{equation}
P_{\rm gal}(k)=b^{2}P(k).
\end{equation}
This approach is the one most frequently adopted in practice, but
the community is becoming increasingly aware of its severe
limitations. A simple parametrisation of this kind simply cannot
hope to describe realistically the relationship between galaxy
formation and environment (Dekel \& Lahav 1999).

\section{Large-scale Structure: Past and Present}

\subsection{Modelling}

Models of structure formation involve many ingredients which
interact in a complicated way: (i) A background cosmology,
basically  a choice of $\Omega_0$, $H_0$ and $\Lambda$ if we are
prepared to stick with the Robertson--Walker metric (1) and the
Einstein equations (3)-(5); (ii) an initial fluctuation spectrum,
usually taken to be a power-law usually with  $n=1$; (iv) a choice
of fluctuation mode, usually adiabatic; (iii) a statistical
distribution of fluctuations usually Gaussian; (v) a transfer
function, which requires knowledge of the relevant proportions of
`hot', `cold' and baryonic material as well as the number of
relativistic particle species; (vi) a `machine' for handling
non-linear evolution, so that the distribution of galaxies and
other structures can be predicted, usually an $N$-body code, an
approximated dynamical calculation or simply, with fingers
crossed, linear theory; (vii) a prescription for relating
fluctuations in mass to fluctuations in light, frequently the
linear bias model. I will now discuss how the attitude to these
ingredients has changed in the past, and is likely to in the near
future.

Historically speaking, the first  model incorporating non-baryonic
dark matter to be seriously considered was the hot dark matter
({\bf HDM}) scenario, in which the universe is dominated by a
massive neutrino with mass around 10--30 eV. This scenario has
fallen into disrepute because the copious free streaming it
produces smooths the matter fluctuations on small scales and means
that galaxies form very late. The favoured alternative for most of
the 1980s was the cold dark matter ({\bf CDM}) model in which the
dark matter particles undergo negligible free streaming owing to
their higher mass or non-thermal behaviour. A `standard' CDM model
({\bf SCDM}) then emerged in which the cosmological parameters
were fixed at $\Omega_0=1$ and $h=0.5$, the spectrum was of the
Harrison--Zel'dovich form with $n=1$ and a significant bias,
$b=1.5$ to $2.5$, was required to fit the observations (Davis et
al. 1985).

The SCDM model was ruled out by a combination of the COBE-inferred
amplitude of primordial density fluctuations, galaxy clustering
power-spectrum estimates on large scales, cluster abundances and
small-scale velocity dispersions (Peacock \& Dodds 1996). It seems
the standard version of this theory simply has a transfer function
with the wrong shape to accommodate all the available data with an
$n=1$ initial spectrum. Nevertheless, because CDM is such a
successful first approximation and seems to have gone a long way
to providing an answer to the puzzle of structure formation, the
response of the community has not been to abandon it entirely, but
to seek ways of relaxing the constituent assumptions in order to
get a better agreement with observations. Various possibilities
have been suggested.

If the total density is reduced to $\Omega_0\simeq 0.3$, which is
favoured by many  arguments, then the size of the horizon at
matter--radiation equivalence increases compared with SCDM and
much more large-scale clustering is generated. . This is called
the open cold dark matter model, or {\bf OCDM} for short. Those
unwilling to dispense with the inflationary predeliction for flat
spatial sections have invoked $\Omega_0=0.2$ and a positive
cosmological constant  to ensure that $k=0$; this can be called
{\bf $\Lambda$CDM} and is also favoured by observations of distant
supernovae . Much the same effect on the power spectrum may also
be obtained in $\Omega=1$ CDM models if matter-radiation
equivalence is delayed, such as by the addition of an additional
relativistic particle species. The resulting models are usually
called {\bf $\tau$CDM}.

Another alternative to SCDM  involves a mixture of hot and cold
dark matter ({\bf CHDM}), having perhaps $\Omega_{\rm hot}=0.3$
for the fractional density contributed by the hot particles. For a
fixed large-scale normalisation, adding a hot component has the
effect of suppressing the power-spectrum amplitude at small
wavelengths . T Another possibility is to invoke non-flat initial
fluctuation spectra, while keeping everything else in SCDM fixed.
The resulting `tilted' models, {\bf TCDM}, usually have $n<1$
power-law spectra for extra large-scale power and, perhaps, a
significant fraction of tensor perturbations. Models have also
been constructed in which non-power-law behaviour is invoked to
produce the required extra power: these are the broken
scale-invariance ({\bf BSI}) models.

\subsection{Past Observational Developments}
In 1986, the CfA survey (de Lapparent, Geller \& Huchra 1986) was
the `state-of-the-art', but this contained redshifts of only
around 2000 galaxies with a maximum recession velocity of $15~000$
km s$^{-1}$. The subsequent Las Campanas survey contained around
six times as many galaxies, and goes out to a velocity of $60~000$
km s$^{-1}$ (Shectman et al. 1996). Quantitative measures of
spatial clustering obtained from these data sets offer the
simplest method of probing $P(k)$, assuming that these objects are
related in some well-defined way to the mass distribution and
this, through the transfer function, is one way of constraining
cosmological parameters. For example, Peacock \& Dodds (1996) made
compilations of power-spectra of different kinds of galaxy and
cluster redshift samples. Within the (considerable) observational
errors, and the uncertainty introduced by modelling of the bias,
all the data lie roughly on the same curve. A consistent picture
thus emerged in which galaxy clustering extends over larger scales
than is expected in the standard CDM scenario. It was difficult to
say much in terms of testing the variations on the CDM theme I
have discussed so far, however, because of the sparseness and
limited scale coverage of the available data.

\subsection{The Present: Entering the Precision Era}

The next generation of redshift surveys, prominent among which are
the Sloan Digital Sky Survey  of about one million galaxy
redshifts (Gunn \& Weinberg 1995) and an Anglo-Australian
collaboration using the two-degree field facility (Colless et al.
2001). The latter survey, called 2dFGRS, has now finished taking
data while the Sloan Survey is still in progress. Both exploit
multi-fibre methods that can obtain 400 galaxy spectra in one go,
and will increase the number of redshifts by about two orders of
magnitude over those previously available. The huge increase in
survey depth (2dFGRS reaches redshifts $z\sim 0.3$) has allowed a
much better measurement of the matter power-spectrum (Percival et
al. 2001) and better statistics have allowed some progress to be
made using higher-order statistical diagnostics of non-linearity
and bias (Verde et al. 2002).

It is evident from Figure 2 that, although the three non-SCDM
models are similar at $z=0$, differences between them are marked
at higher redshift. This suggests the possibility of using
measurements of galaxy clustering at high redshift to distinguish
between models and reality. This has now become possible, with
surveys of galaxies at $z \sim 3$ already being constructed
(Steidel et al. 1998, 1999). Unfortunately, the interpretation of
these new data is less straightforward than one might have
imagined. If the galaxy distribution is biased at $z=0$ then the
bias is expected to grow with $z$ (Davis et al. 1985). If galaxies
are rare peaks now, they should have been even rarer at high $z$.
There are also many distinct possibilities as to how the bias
might evolve with redshift (Matarrese et al. 1997; Moscardini et
al. 1998; Coles et al. 1998).

But large-scale structure is not just about clustering power
spectra. There are other ways in which it is possible to use
information about the velocities of galaxies to constrain models
(Strauss \& Willick 1995). Probably the most useful information
pertains to large-scale motions, as small-scale data populate the
highly nonlinear regime. The basic principle is that velocities
are induced by fluctuations in the total mass, not just the
galaxies. Comparing measured velocities with measured fluctuations
in galaxies with measured fluctuations in galaxy counts, it is
possible to constrain both $\Omega$ and $b$. From equations (10)
to (13) it emerges that
\begin{equation}
{\bf v} = - {2f\over 3\Omega H a}{\bf \nabla_x}\phi + {\mbox{
const}\over a(t)}, \label{eq:l15}
\end{equation}
which demonstrates that the velocity flow associated with the
growing mode in the linear regime is curl-free, as it can be
expressed as the gradient of a scalar potential function. Notice
also that the induced velocity depends on $\Omega$. This is the
basis of a method for estimating $\Omega$ which is known as
POTENT. Since all matter gravitates, not just the luminous
material, there is a hope that methods such as this can break the
degeneracy between clustering induced by gravity and that induced
statistically, by bias. See Dekel (1994) for a review. These
methods are prone to error if there are errors in the velocity
estimates. Perhaps a more robust approach is to use peculiar
motion information indirectly, by the effect they have on the
distribution of galaxies seen in redshift-space (i.e. assuming
total velocity is proportional to distance). The information
gained this way is statistical,  but less prone to systematic
error (Peacock et al. 2001) and the evolution of the effect with
redshift is also a test of cosmological models (Ballinger, Peacock
\& Heavens 1996).

Another class of observations that can help break the degeneracy
between models involves gravitational lensing. The most
spectacular forms of lensing are those producing multiple images
or strong distortions in the form of arcs. These require very
large concentrations of mass and are therefore not so useful for
mapping the structure on large scales. However, there are lensing
effects that are much weaker than the formation of multiple
images. In particular,  distortions producing a shearing of galaxy
images promise much in this regard (Kaiser \& Squires 1993). With
the advent of new large CCD detectors, this should soon be
realised (Mellier 1999).

The combination of lensing, peculiar motions and galaxy clustering
studies would be impressive enough even without the dramatic
arrival of  WMAP on the scene (Bennett et al. 2003). The WMAP data
have really heralded the precision era, allowing direct
determinations of the primordial fluctuation spectrum and the
basic cosmological parameters in a manner that bypasses most of
important sources of uncertainty in clustering analysis.

The bumps and wiggles shown in the transfer functions of Figure 1
do find themselves into the present-day spectrum of galaxy
clustering, but they are strongly affected by non-linear evolution
on the way. Moreover galaxy surveys probe the distribution of
luminous matter so one can't infer the matter spectrum directly
from that of galaxies without a model for the bias. Galaxies also
have peculiar motions so their redshifts do not exactly represent
their proper distances. Survey determinations of $P(k)$ will
inevitably be harder to interpret to those obtained from the
cosmic microwave background, where none of these complications
arise (Hinshaw et al. 2003). This is the reason for the tremendous
precision of WMAP's determination of cosmological parameters
(Spergel et al. 2003), which can be improved still further by
combining constraints from the 2dFGRS and lensing studies.
 Nevertheless, there are very strong
possibilities that a redshift survey performed with the SKA could
 probe both the spectrum and the background cosmology, for
example by using the `wiggles' as standard rulers (e.g. Blake \&
Glazebrook 2003).

\section{The Way Ahead: A Role for SKA Redshift Surveys?}

WMAP, 2dFGRS and the other manifestations of precision cosmology
have certainly made great strides towards the determination of the
cosmological parameters. The standard model that has emerged
(which is very similar to the $\Lambda$CDM model described above).
Although it would be premature to say that no departures from this
model are possible, the emphasis as far as galaxy clustering is
concerned will be away from its use as a probe of the background
cosmology. So what is the future? And is there a role for the SKA
in cosmological studies other than consistency checks of the
standard model? The answer to both questions is emphatically
``yes''.

One can see evidence of a new direction already. Some of the most
interesting results to have emerged from 2dFGRS concern the
clustering of galaxies selected by spectral type (Madgwick et al.
2003). Preliminary results from the Sloan Digital Sky Survey
reveal a complicated dependence of clustering the colours of
selected galaxies (Zehavi et al. 2002). There is evidence of
clustering dependence on intrinsic galaxy properties emerging also
from infra-red selected galaxies (Hawkins et al. 2001). While
these dependencies are simply a nuisance when it comes to
determining cosmological parameters, they indicate that the
large-scale distribution of galaxy clustering may hold clues to
their formation process. The relative clustering strength of
different populations may be complex and scale-dependent,
requiring more sophisticated description that the simple bias
parameters described above.

In principle observations such as these can be used to test
semi-analytic models of galaxy formation of the form discussed by Baugh
et al. (1998). On the other hand, all the classes of galaxy
mentioned are selected by radiation coming from sources with a
complex and poorly understood formation process. The Square
Kilometre Array could produce a great step forward in this area,
by mapping galaxy positions and redshifts in neutral hydrogen via
the 21cm line. Detailed theoretical predictions are so far
lacking, but two ``straw man'' surveys are described in the SKA
science case.

\subsection{The SKA Shallow Survey}

The first case is a ``traditional'' redshift survey along the
lines of 2dFGRS but using 21cm to select the galaxies. Depending
on the eventual choice of instrumental sensitivity, such a survey
might take 12 months, cover about 1000 square degrees of sky and
be capable of detecting galaxies out to $z\sim 2$; compare the
limit $z \sim 0.3$ of 2dFGRS. All in all, this means a survey of
around $\sim 10^{7}$ galaxies in a volume of order $10^7$ Mpc.
This is impressive enough in itself, but such a survey would also
bring with it the possibility of HI Tully-Fisher measurements for
the galaxies in it. In this respect its nearest present relative
is the 6dF galaxy survey described at:

\begin{center} {\tt \begin{verbatim}
http://www.mso.anu.edu.au/6dFGS/6dF_survey_plan.html
\end{verbatim}}
\end{center}

The potential to combine redshifts and Tully-Fisher distances
enables velocity field mapping on an immense scale.

\subsection{The SKA Pencil Beam Survey}

An alternative mode of redshift survey for SKA is to look at a
smller area for much longer. Using the same sensitivity as in the
previous example of a shallow survey, a 360 hour survey covering
one square degree could contain $10^5$ galaxies. A present-day
$L^*$ galaxy could be detected in its HI emission out to a
redshift $z\sim 3$. The limiting HI mass would be a few times
$10^9 M_\odot$ at $z\simeq 4$ and of order $10^8 M_\odot$ at
$z\simeq 1$.

The possibility of detecting objects at high redshift offers the
prospect of constraining models of galaxy formation extremely
strongly. In all hierarchical clustering models, the bias
associated with galaxies increases dramatically with redshift.
This results in a strange conspiracy: the matter correlations
decrease with increase redshift while the bias increases in
compensation, producing a very slow evolution of measured
clustering with epoch. However, the probes we have of
high-redshift clustering, such as Lyman-break galaxies (Steidel et
al. 1998, 1999) and QSOs (Outram et al. 2001), suffer from low
sampling density and uncertain interpretation of the host object.
More importantly, the supply of cold gas plays a central role in
the detailed semi-analytic models of galaxy formation and the
evolution of HI mass function with redshift will be a decisive
test of the basic framework. However, much theoretical work is
needed to make detailed predictions for such surveys. The
hierarchical nature of structure formation involves gas being
distributed in less massive haloes at high redshift, but gas is
also used up to form stars as time goes. The number of HI sources
seen as a function of redshift may be drastically different from
that expected of a non-evolving population of present-day
galaxies.

\section{Discussion and Conclusions}

I have emphasized the importance of clustering properties and
their implications for galaxy and large-scale structure formation.
There will no doubt be many that disagree with this emphasis.
Large-scale matter power spectrum determination will be possible
using SKA and will be enormously better even that 2dFGRS or Sloan.
Such studies are well-worth doing, as are the numerous possible
tests of departures from the standard model, especially with
respect to the possible forms of dark energy such as quintessence
(Huterer \& Turner 2001). A 21cm survey would be better fitted to
such a task than QSO surveys (e.g. Outram et al. 2001) because of
the higher sampling density.

Interesting though the results of such studies will be, they will
almost certainly turn out merely to provide consistency checks on
a cosmology largely fixed by studies of the cosmic microwave
background. For me, the  the distribution of cold gas on
large-scales, how it relates to stellar populations of various
kinds and how the supply of this gas has evolved with cosmic epoch
offers the richest scientific possibilities. What is now needed is
proper theoretical modelling of HI-selected galaxies to produce
mock catalogues to drive the science case further forward. Watch
this space.

\acknowledgements

This style file is based on one provided by PASP. I thank the
editors for their patience!

\end{document}